\documentclass[twoside]{article}
\usepackage{graphicx} 
\usepackage{subfigure}
\usepackage{xspace}
\usepackage{url}
\usepackage{algorithm}
\usepackage{algorithmic}
\usepackage{times}
\usepackage{amsfonts}
\usepackage{amsmath}
\usepackage{fancyhdr}

\begin{document}

\newcommand{\ieee}{{\sc ieee} 802.11\xspace}
\newcommand{\lp}{{\sc lp}\xspace}
\newcommand{\milp}{{\sc milp}\xspace}
\newcommand{\wsnet}{WsNet\xspace}
\newcommand{\rwcds}{r-WCDS\xspace}
\newcommand{\potatoes}{\texttt{potatoes}\xspace}

\newcommand{\T}{\mathcal{T}}
\newcommand{\CT}{\mathcal{CT}}
\newcommand{\leader}{\mathcal{L}eader}
\newcommand{\cluster}{\mathcal{C}luster}
\newcommand{\rootleader}{\mathcal{R}oot\mathcal{L}eader}
\newcommand{\clusterleader}{\mathcal{C}luster\mathcal{L}eader}
\newcommand{\rootcluster}{\mathcal{R}oot\mathcal{C}luster}

\newcommand{\parent}{\mathcal{P}arent}
\newcommand{\children}{\mathcal{C}hildren}
\newcommand{\clusterLinks}{\mathcal{C}luster\mathcal{L}inks}
\newcommand{\clusterRoles}{\mathcal{C}luster\mathcal{R}oles}

\newcommand{\rootleaderdistance}{\mathcal{R}oot\mathcal{L}eader\mathcal{D}istance}
\newcommand{\leaderDistance}{\mathcal{L}eader\mathcal{D}istance}

\newcommand{\V}{\mathcal{V}}
\newcommand{\E}{\mathcal{E}}
\newcommand{\G}{\mathcal{G}}
\newcommand{\W}{\mathcal{W}}
\newcommand{\capa}{BW}
\newcommand{\NbCh}{nbCH}  
\newcommand{\ch}{CH}  
\newcommand{\Traffic}{T}   
\newcommand{\TrafficMin}{T_{min}}   
\newcommand{\role}{role}

\title{A Divide-and-Conquer Scheme for Assigning Roles in Multi-Channel Wireless Mesh Networks}
\author{
Beno\^it Darties, Fabrice Theoleyre, and Andrzej Duda
\vspace{0.4cm}\\
{\small Grenoble Informatics Laboratory}
\\
{\small CNRS and Grenoble-INP}
\\
{\small 681 rue de la passerelle, BP72}
\\
{\small 38402 Saint Martin d'Heres, France}
\\
{\small Email: \url{{firstname.lastname}@imag.fr}}
}

\date{}

\maketitle

\begin{abstract}
A multi-channel MAC seems to be an interesting approach for improving network throughput by multiplexing transmissions over orthogonal channels. In particular, Molecular MAC has recently proposed to modify the standard \ieee DCF access method to use dynamic channel switching for efficient packet forwarding over multiple hops. However, this MAC layer requires role and channel assignment to nodes: some of them use a static channel, while others dynamically switch to neighbor channels on-demand. To assign roles and channels, we extend the notion of the Weakly Connected Dominating Set, the structure already used in clustering. More precisely, we adapt the WCDS structure and introduce new constraints to define what we call a \emph{reversible WCDS} (\emph{r-WCDS}), which is particularly suitable for wireless mesh networks operating under Molecular MAC. We propose a divide-and-conquer scheme that partitions the network into clusters with one leader per cluster solving a \milp formulation to assign roles in its cluster. By appropriately defining the roles at the border of clusters, we maintain global connectivity in the \rwcds. Finally, our simulations show that the performance of the propose scheme is close to a centralized algorithm.
\end{abstract} 




\section{Introduction}

We consider wireless mesh networks composed of a large number of wireless routers providing connectivity to mobile nodes. They begin to emerge in some regions to provide cheap network connectivity to a community of end users. Usually they grow in a {\em spontaneous} way when users or operators add more routers to increase capacity and coverage.

We assume that mesh routers benefit from sufficient resources, may
only move, quit, or join occasionally, so that the topology of a
typical mesh network stays fairly stable.  The organization of mesh
networks needs to be {\em autonomic}, because unlike the current
Internet, they cannot rely on highly skilled personnel for
configuring, connecting, and running mesh routers. Thus, proposed
protocols must be distributed and self-stabilizing.

One way of organizing the structure of a wireless mesh network is to
construct a Weakly Connected Dominating Set (WCDS) \cite{dunbar97},
which is a well-known clustering scheme in wireless multihop
networks. By appropriately electing clusterheads (called
\emph{dominators} in the WCDS terminology), each node in the network
is a member of one cluster and the network structure results in a
limited number of hops between clusterheads. Such clustering is useful
for limiting the overhead of flooding, introducing a routing
hierarchy, distributing keys \cite{pathan06}, or other network-wide
operations.

To improve network capacity, wireless mesh networks can use multiple
radio channels: multiplexing transmissions on orthogonal channels
allows for parallel transmissions through spectrum spatial reuse and
reduces collisions. Mesh routers can take advantage of parallel
transmissions over neighbor links by using multiple channels at
various time scales. When nodes have multiple interfaces, they can
statically allocate channels to achieve high spatial reuse and good
performance~\cite{Raniwala04, Raniwala05, Alichery05,
  Kodialam05}. Nodes with single interfaces can also benefit from
multiple channels by switching channels on a per frame
basis~\cite{so04,bahl04}. Molecular MAC \cite{molecular09rr} has
recently proposed to modify the standard \ieee DCF access method to
use dynamic channel switching for efficient packet forwarding over
multiple hops. It solves the deafness problem inherent to
multi-channel schemes by assigning a static channel for one part of
nodes ({\em nuclei} of spatially distinct {\em atoms}) and letting
other nodes ({\em electrons}) dynamically switch between
channels. Electrons always initiate transmissions, while nuclei notify
other nodes about pending packets. Molecular MAC outperforms classical
strategies like MMAC~\cite{so04} with respect to throughput, fairness,
and end-to-end delay. However, the authors have left the problem of
assigning roles (nucleus or electron) and channels for future
work.

In this paper, we propose a protocol for organizing a wireless mesh
network according to a suitable structure associated with Molecular
MAC. We adapt the well-known WCDS structure and introduce new
constraints to define what we call a \emph{reversible WCDS}
(\emph{r-WCDS}), which is particularly suitable for wireless mesh
networks operating under Molecular MAC.

The contribution of this paper is threefold:
\begin{itemize}
	\item it provides a formal definition of the reversible WCDS,
	\item we propose a new divide-and-conquer scheme for constructing such a r-WCDS in a distributed way, 
	\item we evaluate the performances of the proposed scheme and compare it with other approaches.
\end{itemize}

This paper is organized as follows. First, we introduce some notations
and define the problem of constructing a WCDS for Molecular MAC. In
Section~\ref{section:milp} we use a Mixed Integer Linear
Programming formulation (\milp) to find a r-WCDS. Then, we present in
Section~\ref{section:potatoes_description} \potatoes, a
divide-and-conquer scheme for constructing a r-WCDS in a scalable
way. We report in Section~\ref{section:performance_evaluation} on the
simulation based performance evaluation of the proposed scheme and
we discuss the related work in Section~\ref{section:related_work}. We
finally conclude the paper and give some perspectives.


\section{Problem formulation and notation}
\label{section:problem_formulation}

We model the network as an undirected graph $G=(\V,\E)$ in which vertices
$\V(\G)$ are the set of nodes and edges $\E(\G)$ are all pairs of nodes
able to directly communicate. We adopt the following classical
notation:
\begin{itemize}
	\item $n=|\V|$ defines the number of nodes in the mesh network,
	\item $N(u)$ is the set of neighbors of $u$ with
          cardinality $\Delta(u) = |N(u)|$,
	\item $\{u,v\}$ denotes the edge between vertices $u$ and $v$, i.e. $\{u,v\} \in \E$,
	\item $\capa$ denotes channel capacity,
	\item $\ch$ is the set of all available channels and $\NbCh =
          |\ch|$ (\ieee{a} provides for instance $12$ orthogonal
          channels in the US).
\end{itemize}

\subsection{Reversible Weakly Connected Dominating Set Problem}

The Weakly Connected Dominating Set (WCDS) \cite{dunbar97} is a widely
used structure in wireless multihop networks.
Formally a WCDS is defined by the set $D\subseteq \V$ such that:
\setlength{\arraycolsep}{0.0em}
\begin{eqnarray}
\forall u \in \{\V-D\} 		& \quad 	,	& \quad \exists v \in D |  v \in N(u) \\
\G=(\V,\E')\ connected 	& \quad 	|	& \E'=\big\{ \{u,v\},  u \in D, v \in \V\big\}
\label{eq:wcds}
\end{eqnarray}
A node of $D$ is often called \emph{dominator} while nodes in  $\V-D$
are called \emph{dominatees}. 

We define the \emph{reversible WCDS} (denoted \rwcds in the rest of the paper) as follows: we only keep the
edges $(dominator,dominatee)$, i.e. edges between dominators are
removed. We will further see that Molecular MAC requires such a
structure. Formally, we transform the second property of
Eq.~\ref{eq:wcds} into:
\begin{equation}
	\G=(\V,\E')\ connected | \E'=\big\{ \{u,v\},  u  \in D, v \in \{\V-D\}\big\}
\label{eq:reversible_wcds}
\end{equation}


\subsection{Relation to Molecular MAC}

We are interested in the reversible WCDS, because Molecular MAC
\cite{molecular09rr} requires such a structure.  Molecular MAC divides
a wireless mesh network into spatially distributed \emph{atoms} so that each
atom uses a fixed channel different from its neighbors. An atom is
composed of a \emph{nucleus} and \emph{electrons}. A nucleus chooses a
channel for its atom and sticks to this channel all the time.  Nodes
at the border of atoms have the role of electrons bonding neighboring
atoms: they forward traffic between atoms by dynamically switching
their channel to communicate with neighboring nuclei.

We can compare Molecular MAC to an extended WLAN: an atom corresponds
to a WLAN cell and a nucleus is a \emph{virtual Access Point} that
interconnects other nodes in the cell (electrons). In mesh
networks, a node needs to communicate with several
cells. Since a node should not miss packets from its virtual AP because it is transmitting packets over another WLAN, Molecular MAC
implements a specific MAC mechanism: a node pulls its data frames
from the AP that buffers frames and transmits the list of pending
destinations in beacons.


\begin{figure}
\begin{center}
	\includegraphics[width=8cm]{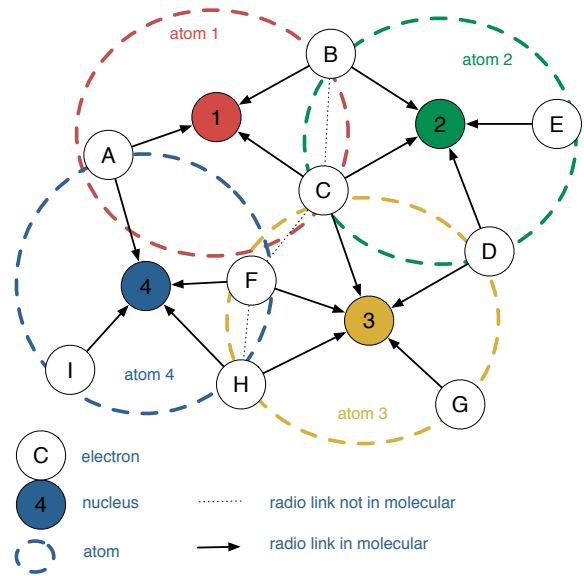}
	\caption{Example molecular topology}
	\label{fig:molecular_topology}
\end{center}
\end{figure}

In Molecular Mesh, we need to assign a role to each node (a nucleus or
an electron) so that the resulting network has the following
properties:
\begin{enumerate}
	\item a node can communicate with any other node via multi-hop forwarding;
	\item only nuclei and electrons can communicate with each
          other, i.e. there is no direct communication between two electrons
          or two nuclei. Indeed, two nuclei do not hear each other,
          because they use different channels. Similarly, two
          electrons continually switch their channels and may suffer
          from deafness;
	\item the capacity of the network should be maximal. In particular, two neighboring atoms, which can interfere, need to use different channels.
\end{enumerate}

Figure~\ref{fig:molecular_topology} illustrates the molecular
organization. The identifiers of nuclei are numbers while they are
letters for electrons. We can note that links between two electrons
are not used, however we need to keep the number of unused links small
to allow for redundant paths in the network for better connectivity
and failure tolerance. 
Clearly, the constraints for assigning roles to nodes lead to a
reversible-WCDS: each dominator corresponds to a nucleus and an
electron to a dominatee.

Constructing the molecular structure also requires that each nucleus
node selects a fixed channel. Simultaneous assignment of roles and
channels so to maximize network capacity is a difficult
problem. However, a good tradeoff between computation time and network
performance consists first of determining the role of a node to obtain
a connected component regardless of channel usage. Once a node has
become a nucleus, it can choose a channel according to a greedy
approach by scanning all available channels and choosing the one with
the minimum load.

\section{MILP Formulation}
\label{section:milp}

We use a \milp(Mixed Integer Linear Programming) formulation already
defined in our previous work \cite{molecular_milp09rr} to assign roles
(i.e. finding a r-WCDS) while maximizing the network capacity.  We
define all the constraints corresponding to flow conservation,
connectivity requirements, throughput maximization as linear
constraints. Its solution leads to the optimal assignment of roles
(nucleus or electron) and channels in a spontaneous mesh network.  We
summarize it below.
\begin{itemize}
\item We assign a role to each node $u\in \V$ represented by variable
  $\role(u) \in \{0,1\}$ with value $1$ if $u$ is a dominator and $0$
  otherwise.
\item Our performance objective is to maximize the global network
  throughput. We assume any-to-any traffic pattern: each node
  communicates with each other node thus giving $n(n-1)$ multihop
  flows. We maximize $\TrafficMin$, the minimum throughput allocated
  to each flow.
\item $\Traffic(u,v,d)$ corresponds to traffic transmitted by $u$
  through link $\{u,v\}$ to destination $d$ for each triplet $(u,v,d)
  | \{u,v\} \in E, d \in V$. Note that $\forall \{u,v\} \in \E,\ T(u,v,u)=0$ (i.e. a node does not generate traffic for itself);
\item $\Delta(u)$ denotes the degree of node $u$, i.e. the number of
  neighbors in the network.
\end{itemize}

\subsubsection{Links between nodes} 
We can only use a link if and only if its endpoints have different roles. Its capacity (the sum of $\Traffic(u,v,d)$ over all
destinations $d$) is zero if both endpoints are dominators (Eq.~\ref{eq:linknucleuselectron_1}) or dominatees (Eq.~\ref{eq:linknucleuselectron_2}):
\setlength{\arraycolsep}{0.0em}

\begin{multline}
	\forall \{u,v\}\in \E, \role(u)+\role(v) \\+  \frac{1}{\capa}\sum_{d\in V}\big(\Traffic(u,v,d)+\Traffic(v,u,d)\big) \leq 2 	
	\label{eq:linknucleuselectron_1}
\end{multline}

\begin{multline}
	\forall \{u,v\}\in \E, \frac{1}{\capa}\sum_{d\in V}\big(\Traffic(u,v,d) + \Traffic(v,u,d)\big) \\ \leq \role(u)+\role(v)
	\label{eq:linknucleuselectron_2}
\end{multline}

\subsubsection{Flow conservation}
Eq. \ref{eq:flowconserv_1} and \ref{eq:flowconserv_2} express the flow
conservation law: the sum of traffic to $d$ forwarded by $u$ is equal to
the sum of traffic to $d$ entering in $u$ and the traffic generated by
$u$ to $d$ (which must be at least equal to throughput
$\TrafficMin$). Eq.~\ref{eq:flowconserv_2} represents the fact that a
destination node must receive exactly $(n-1).\TrafficMin$ total
traffic units ($\TrafficMin$ for each $(n-1)$ sources):

\begin{multline}
	\forall u,d\in \V, d \neq u, \\  \sum_{v\in N(u)} \Traffic(u,v,d) =   \sum_{v\in N(u)} \Traffic(v,u,d)  +  \TrafficMin\label{eq:flowconserv_1} 
\end{multline}
\begin{equation}
	\forall u\in \V,\quad 	\sum_{v\in N(u)}  (n-1).\TrafficMin = \Traffic(v,u,u) \label{eq:flowconserv_2}
\end{equation}

\subsubsection{Capacity of an atom}
All links belonging to an atom share its bandwidth $\capa$:
\begin{equation}
\forall u\in \V,	\quad	\sum_{v\in N(u)}\sum_{d\in V}\big(\Traffic(u,v,d)+\Traffic(v,u,d)\big)   \leq  \capa \label{eq:atomcapa} 
\end{equation}
The constraints are obvious if $u$ is a dominator. If $u$ is a
dominatee, it cannot receive more than \capa, even if it is adjacent
to several dominators because of time sharing mechanisms for switching
between channels.

\subsubsection{Improvement}
Optional inequalities (Eq. \ref{eq:singlecut}) accelerate the \milp
resolution by stating that each dominator is adjacent to at least one
dominatee and reciprocally:
\begin{equation}
\forall u\in \V, 	\quad	1 \leq  \role(u) + \sum_{v\in
  N(u)}\role(v) \leq \Delta(u) \label{eq:singlecut}
\end{equation}

Solving the \milp is computationally expensive for large networks: a
couple of hours is required to obtain the optimal assignment in a mesh
network of $40$ nodes.


\section{\potatoes: a divide and conquer scheme}
\label{section:potatoes_description}

In this paper, we propose a divide-and-conquer scheme: we divide the
network into \emph{clusters} with one \emph{leader} per cluster that
solves the \milp formulation for its cluster. The small size of
clusters leads to good efficiency of obtaining the local \milp
solution. However, we need to enforce additional constraints so that
the union of multiple local r-WCDSs results in a global connected
r-WCDS.  To obtain this goal, we have to define clusters in a certain
manner and fix the roles of some node. We provide below the details of
the mechanisms for constructing the clusters and achieving the global
\rwcds.

\subsection{Approach}

\begin{figure}[!h]
\begin{center}
	\includegraphics[width=10cm ]{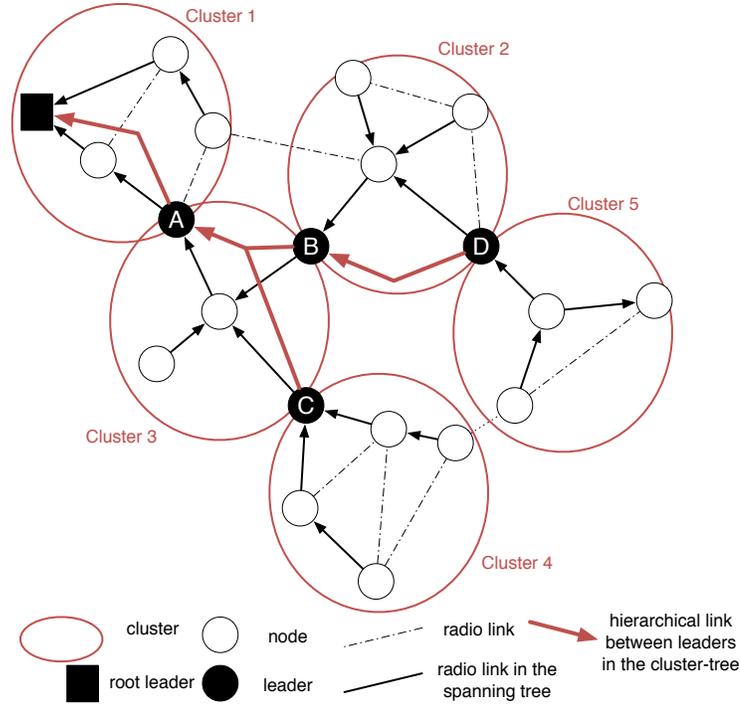}
	\caption{An example cluster-tree.}
	\label{fig:cluster_tree} 
\end{center}
\end{figure}

We propose here a distributed scheme to construct a \rwcds. We first
construct a \emph{rooted cluster tree}, i.e. a tree in which
vertices are clusters and a link exists between clusters if and only if they
share a node. We assume that the network is connected: if several
components exist, the algorithm will be executed independently in each of
them. The \emph{rooted cluster tree} supports distributed role
assignment: one leader per cluster computes the optimal local
assignment in its cluster (i.e. not taking into account other nodes
and radio links). However, we need to limit the dependence between two
clusters, i.e. a node should receive its role from only one
leader. Thus, we add the following constraints with respect to
classical cluster-trees:
\begin{enumerate}
\item One node belongs to at most two clusters. A node that
  belongs to exactly two clusters is a cluster member of the cluster
  higher in the tree and the leader in the other one;
\item the intersection of any two clusters contains at most one node.
\end{enumerate}
Figure~\ref{fig:cluster_tree} presents an example of clustering: there
are five clusters forming a tree hierarchy. We can verify that each
leader belongs to exactly two clusters except the \emph{root leader}
at the root of the cluster-tree: it is the network leader.

The leader of each cluster learns the topology of its cluster,
i.e. node ids and links between nodes. Then, it computes the local
optimal solution with the \milp formulation: all the constraints are
translated into linear inequalities and the global objective consists of
maximizing the throughput. After computing the roles for its cluster,
the leader has just to notify its cluster members about their
roles. 

The assignment is optimal in one cluster, but the union of local
assignments does not necessarily leads to a global optimal r-WCDS, because
leaders find the optimal role assignment inside clusters and not among
clusters. 
We need to enforce that network nodes belonging to different clusters
end up with the same role in the global r-WCDS. To achieve this goal,
we use the hierarchy of the cluster-tree and proceed in the following
way:
\begin{enumerate}
	\item the leader assigns a role to all its cluster members,
	\item a node that belongs to one cluster just uses the
          assigned role,
	\item a node that belongs to more than one cluster is the
          leader of a cluster down in the hierarchy (e.g. node $C$ in
          Figure~\ref{fig:cluster_tree}) and it will receive its role
          from the leader upper in the hierarchy (e.g. node $A$).
\end{enumerate}

Finally, at least one path exists in the rooted cluster-tree that uses the hierarchy of clusters. We will now define more formally this algorithm. 

\subsection{Notation}
We will use the following definitions:

\begin{itemize}
\item a cluster is a connected subgraph $\cluster$ of network graph
  $\G$ induced by the subset of nodes $S\subseteq \V(\G)$ with
  $\V(\cluster)=S$ and $\E(\cluster) = \big\{\{u,v\}| u,v\in \V(\G) \
  and\ \{u,v\}\in \E(\G)\big\}$;
	
	\item $\T$ denotes a spanning tree used to build the cluster-tree;
	
	\item $\CT$ represents the cluster-tree structure; 
	
	\item network leader node denoted by $\rootleader$ is the
          root of spanning tree $\T$. The unique cluster containing
          $\rootleader$ is denoted $\rootcluster$ and is the root of
          $\CT$;
	
	\item $V(\cluster)$ denotes the set of nodes belonging to $\cluster$;
		
	\item $\leader(\cluster)$ denotes the leader of $\cluster$. Obviously, $\rootleader$ is the leader of $\rootcluster$. If $\cluster \neq \rootcluster$, $\clusterleader$ is the node belonging to $\cluster$ and to the upper cluster in the cluster-tree. Formally, $leader(\cluster) = \V(\cluster) \cap V(parent(\cluster))$ with $parent(\cluster)$ being the upper cluster in $\CT$;
	
	\item intra-cluster edges are links between two nodes of the
          same cluster. Any other link is an inter-cluster edge.
	
	\item $role(u)$ denotes the role of node $u$ (either dominator
          or dominatee). 
\end{itemize}

\subsection{Cluster-Tree Construction}

We assume that nodes periodically broadcast
\texttt{hello} messages to discover their neighbors. Usually a network
uses this kind of protocols for supporting other functions
such as routing---we can just piggyback additional information
required by \potatoes in such messages. 

According to the definitions above, all leader nodes except
$\rootleader$ belong to exactly two clusters. In the example presented
in Figure \ref{fig:cluster_tree}, $\rootcluster$ is located on the
top-left corner and the rooted cluster tree has five clusters and the
depth of two. The example tree-cluster has two inter-cluster edges.
As our algorithm finds the optimal role assignment inside clusters,
the presence of the intra-cluster edges in the final \rwcds will
depend on the role assigned to each node at the border.

The tree-cluster construction proceeds as follows.
\begin{enumerate}
\item The network elects $\rootleader$.




\item Nodes construct a spanning-tree rooted at the network leader
  (i.e. $\rootleader$). A node propagates the minimum id received by
  neighbors (or its own id if it is lower) by piggybacking the tuple
  $<min\_id,distance,seq\_nb>$ in \texttt{hello} messages. A node
  updates its distance to $\rootleader$ only if the \texttt{hello}
  message contains a larger $seq\_{nb}$ than the current
  $seq\_{nb}$. Only $\rootleader$ increments the
  sequence number in each \texttt{hello}: a node can safely decide that $\rootleader$ has
  left the network, if the sequence number is not incremented during
  $T_{dead}$ time interval.  A node also includes in its
  \texttt{hello} messages the id of its parent in the tree. In this
  way, any node can maintain the list of its children in $\T$.

\item We define clusters and their leaders based on their position in
  the spanning tree. A new cluster is created when the distance to the
  leader upper in the spanning tree is exactly $D+1$ hops:
	\begin{enumerate}
		\item $\rootleader$ becomes the first leader; 
		\item each node piggybacks the identity of its leader,
                  $\leader(u)$ in \texttt{hello} messages. It becomes
                  the first leader on the path in $\T$ towards
                  $\rootleader$;
		\item the tree is divided so that each node is at most
                  $D$ hops away from its leader and the leader has the
                  minimum depth in the tree among all its cluster
                  members. \potatoes elects the nodes that are
                  exactly $\equiv 0\ [D]$ hops away from the
                  $\rootleader$ ($[D]$ stands for \emph{modulo D}).
	\end{enumerate}
      \item a cluster is finally defined for each leader
        $\mathcal{L}$ and contains all the nodes with
        $\leader(u)=\mathcal{L}$.
\end{enumerate}

Let us consider the topology in Figure~\ref{fig:cluster_tree} with a
cluster radius of 2. The figure presents the spanning tree used for
the cluster-tree construction: each node is at most 2 hops in the
spanning tree away from its leader. Border leaders maintain hierarchical
relationships: $C$, the leader of cluster $4$ is one of the children
of leader $A$ in cluster $3$.

\subsection{Learning Cluster Topology}

A leader should know the internal topology of its cluster before
solving the \milp optimization. A node could flood the list of its
neighbors in the cluster, but this implies high overhead. We can note
that only the leader needs to know the topology of the whole
cluster. Thus, we can efficiently use tree $\T$ to propagate the
topology information and merge it along the tree.

At the beginning, each node discovers new neighbors through \texttt{hello} messages. Besides, a node maintains a local \emph{topology table} that contains
all links for which one extremity is a child in the
spanning tree $\T$. This topology table is restricted to the
children in the same cluster and is recursively fed: each node
piggybacks its topology table in each \texttt{hello} and
updates it with the information received from its children (their
topology table and their list of neighbors). Obviously, the
table of the leader contains the global vision of the cluster
topology, while the topology tables of other nodes contain only
partial information.

To be fault-tolerant, a node updates its topology table to take
into account joining or leaving neighbors. Moreover, each entry of the
local table contains a \texttt{child-id} field. Each time a child
receives a \texttt{hello} message, it flushes the entries in the local
topology table and replaces them by the new ones. In the same way, a
child that has changed its parent is simply removed from each local
topology table.

\subsection{Role Assignment}
\label{subsection-role-assignment}

If its topology table remains unchanged for a sufficient time, each leader learns the topology of its cluster, computes the local optimal assignment, and sends it to the cluster members.


We force each leader to have a predefined role by adding the
following constraints in the \milp optimization:

\begin{itemize}
\item $\rootleader$ becomes the first dominator.
	
\item If the cluster radius ($D$) is even, each leader can safely take
  the role of a dominator. Each leader $\mathcal{L}$ (except
  $\rootleader$) belongs to exactly two clusters: it is the leader of
  $\cluster$ and a member of $Parent(\cluster)$. Since the radius is
  even, a path alternating dominatees and dominators can link
  $\mathcal{L}$ to the leader in $Parent(\cluster)$;

\item If the cluster radius is odd, the leader $\mathcal{L}$ should
  have the inverse role\footnote{roles are either dominator or
    dominatee.\label{footnote:inverse_roles}} of its leader in
  $Parent(\cluster)$. By alternating dominatees and dominators, we
  would obtain a valid path;
\end{itemize}

Each leader $\mathcal{L}$ in $\cluster$ receives its role from
$Parent(\cluster)$. This role is fixed in the \milp formulation by
adding the linear constraint $role(\mathcal{L})=dominator$ or
$role(\mathcal{L})=dominatee$ according to the role given to
$\mathcal{L}$. Thus, \milp resolution can be fully distributed in each
cluster: $L$ does not have to wait for the role assigned by the leader
of $Parent(\cluster)$ and can directly compute roles in $\cluster$
once it knows the topology.

Let consider the topology in Figure~\ref{fig:cluster_tree}. We can
observe that the \milp formulation can find a solution in a cluster if
we fix the role of all the leaders to dominators since the cluster
radius is even. For instance, leader $A$ in cluster $3$ will execute
its \milp with the role of $A$, $B$, and $C$ already fixed to
$dominator$. In particular, we can in particular ``color'' all the
nodes with an even depth as dominators and others as dominatees in the
spanning tree $\T$. Clearly, a path alternating dominators and
dominatees can interconnect any pair of leaders in the same cluster
leading thus to an achievable solution. In this way, we guarantee that
dominators globally form a valid \rwcds.

This simple optimization accelerates convergence and results in good
performances. Moreover, we keep at least one connected solution in
each cluster that consists for each node to alternate roles between
nuclei and electron according to the distance to its leader in the
cluster.

After solving the \milp optimization $\leader$ sends the list of its
cluster members and their assigned roles along the tree.

\subsection{Discussion}

The presented scheme locally optimizes role assignment. It presents the following advantages:
\begin{enumerate}
\item by splitting the network into clusters, the algorithm is
  scalable and succeeds to find a suitable solution with a reasonable
  computational cost;
\item it is distributed, because it relies on the distributed
  construction of the cluster-tree and role assignment;
\item it is fault-tolerant by taking into account joining and leaving
  nodes as well as lost \texttt{hello} messages.
\end{enumerate}


\section{Performance evaluation}
\label{section:performance_evaluation}

We have simulated the proposed protocol in \wsnet\cite{wsnet} using
the COIN-CBC linear programming library \cite{coin-cbc}. We randomly
place nodes in a simulation area. Nodes use the \ieee{a} network
interface to communicate with each other with the radio range of $10$
units and the interference range of $30$ units. By default, the mesh
network is composed of $50$ nodes with on the average $10$ neighbors. We
adjust the simulation area to obtain given density.

The results correspond to statistics averaged over $10$ different
simulations of $240$ seconds. The graphs present averaged values with
$95\%$ confidence intervals. We compare the performance of the
centralized \milp formulation (OPT), \potatoes, the Maximum
Independent Set protocol (MIS) and a self-stabilizing Spanning Tree
(ST) (\cite{molecular_milp09rr}). We measure the following performance
indices:
\begin{itemize}
\item minimum throughput $T_{min}$: the minimum throughput guaranteed
  for each flow extracted from the \milp formulation. We consider the
  normalized channel capacity (i.e. $\capa = 1$);
\item average route stretch factor: the average ratio of the length of
  the shortest route through the \rwcds and the length of the
  shortest route in the original graph.
\end{itemize}

\begin{figure}[!h]
\centering
	\includegraphics[width=8cm]{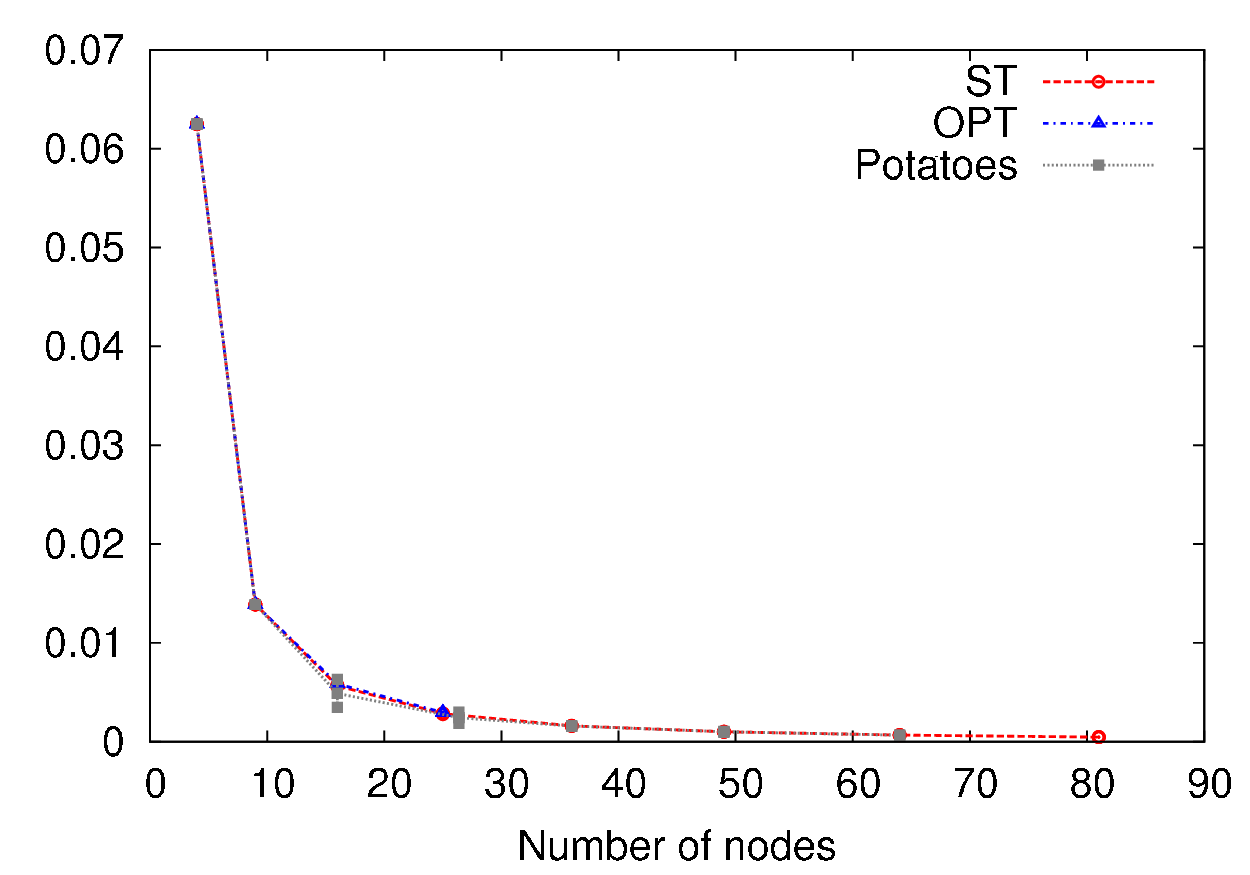}        
	\caption{Minimum throughput $\TrafficMin$ for a varying number
          of nodes in a grid}
\label{fig:grid-min_flow}
\end{figure}

\subsection{Grid Topology: Capacity}

We have first compared the performance of the different protocols in a
grid topology: nodes are placed regularly in a squared grid, the
length of each cell in the grid being the radio range. Such a grid can
represent a regular mesh network deployed for instance by a
telecommunication operator. We do not report the results for MIS since
it leads to a disconnected network in most cases.

We have measured $\TrafficMin$, the minimum throughput allocated to
each flow obtained with the \milp formulation for a varying number of
nodes (cf. Figure~\ref{fig:grid-min_flow}). We can observe that all
the protocols perform quite similarly. In particular, the spanning
tree strategy achieves to find optimal roles and channels: in a grid,
the pruning strategy is inefficient since the number of neighbors
is limited. Thus, ST consists of marking as dominators all the nodes
with an even depth (on average a half of the nodes in a random
spanning tree) and as dominatees other nodes.

\begin{figure}[!h]
\centering
	\includegraphics[width=8cm]{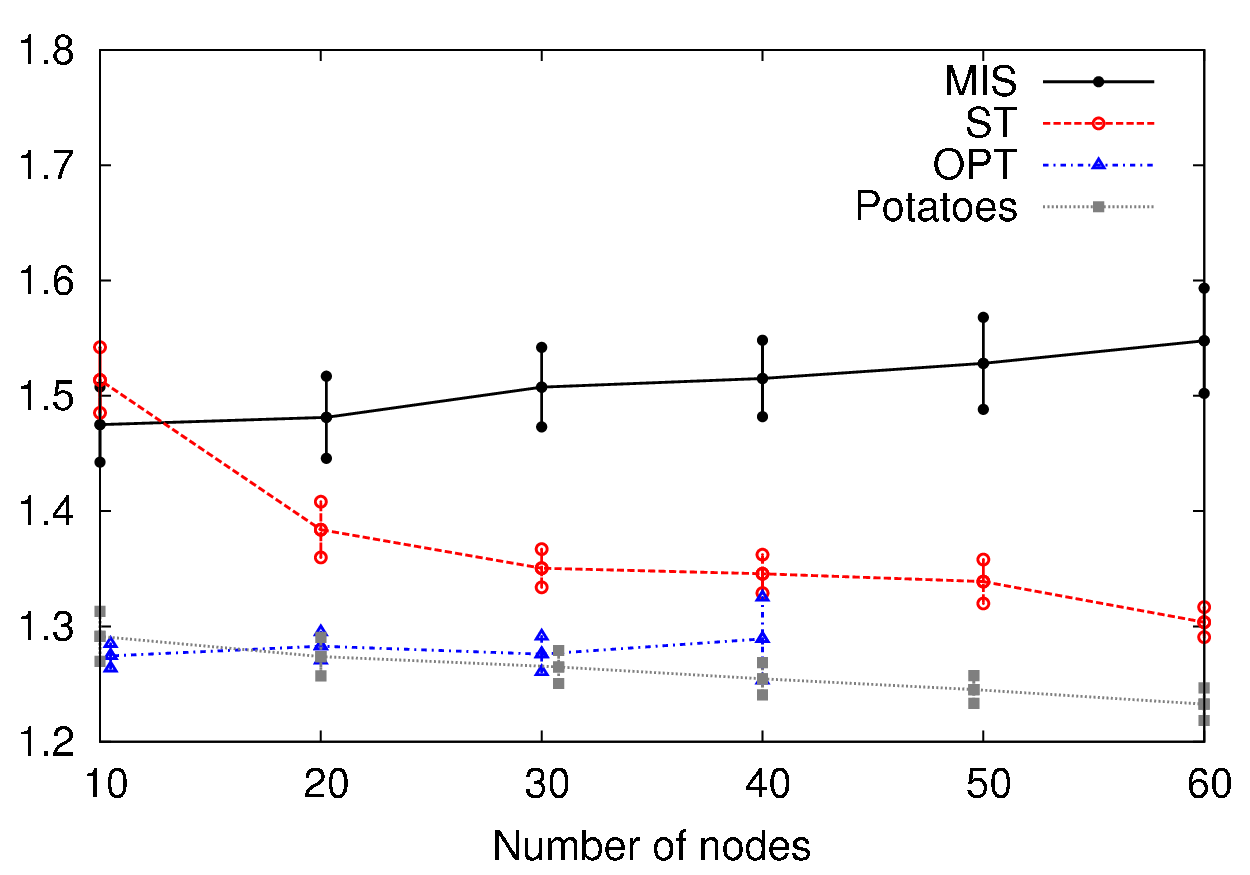}        
	\caption{Average route stretch factor for a varying number of
          nodes in a random network.}
\label{fig:nb_nodes-route_stretch_factor}
\end{figure}

\subsection{Random Topology: Route Stretch Factor}

Then we have considered a random topology of a given density ($10$ neighbors
on the average). We measure the {\em route stretch factor}: the ratio
of the route length in a molecular mesh and in the original graph
(cf. Fig.~\ref{fig:nb_nodes-route_stretch_factor}). A stretch factor
of $1$ means that only the shortest routes are used. For MIS, we
discard isolated nodes since the stretch factor would become infinite
in this case. Thus, we tend to under-estimate the real stretch factor
for MIS.  OPT results in an average stretch
factor around $1.3$ for any number of nodes. However, OPT is not
scalable---results become difficult to obtain in a reasonable time for
more than $40$ nodes (50 nodes require more than 3 hours) and almost
impossible to be obtained for more than $60$ nodes. This explains why
we do not plot results for $50$ and $60$ nodes under OPT in
Figure~\ref{fig:nb_nodes-route_stretch_factor}.

The performance of \potatoes and OPT is very similar. \potatoes
achieves to construct a \rwcds with a maximum number of radio links:
the routes are often the shortest ones. Moreover, \potatoes is much
more scalable than OPT and achieves to compute distributively a \rwcds
even for larger networks. This shows that the divide-and-conquer approach
is suitable for our problem.

ST uses longer routes, but the stretch factor tends to decrease when
the number of nodes increases. The stretch factor for MIS is large and
a flow consumes more bandwidth since it is relayed by more nodes on
average.

\begin{figure}[!h]
\centering
	\includegraphics[width=8cm]{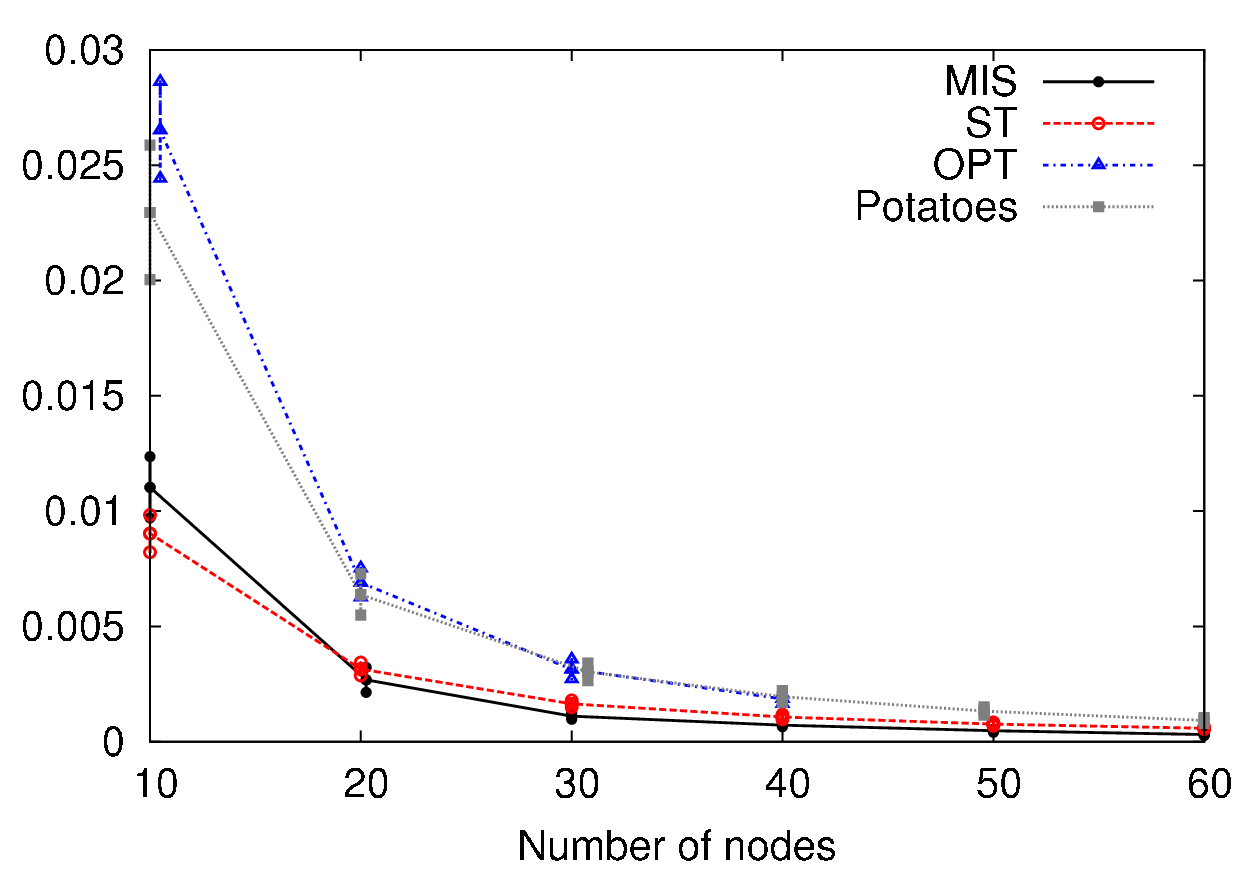}        
	\caption{Minimum throughput $\TrafficMin$ for a varying number
          of nodes in a random network.} 
\label{fig:nb_nodes-min_flow}
\end{figure}

\subsection{Random Topology:  Capacity}

Finally, we have measured the minimum throughput $T_{min}$ in a random
topology (cf. Figure~\ref{fig:nb_nodes-min_flow}). This metric
corresponds to the throughput we can obtain with Molecular MAC.

Obviously, the OPT protocol gives us an upper bound since it finds
role assignment maximizing the objective. However, scalability issues
do not allow to obtain enough significant results when we have more
than $40$ nodes. Here again \potatoes results are very close to those
of OPT.

We can note that MIS and ST achieve much lower throughput: they do not
succeed in maximizing the number of links remaining in the \rwcds
topology, which leads to a lower throughput.  For the density we have used,
the average performance of both strategies is less than half of the
performance obtained with \potatoes.


\section{Related work}
\label{section:related_work}

Clustering creates groups of nodes, which is particularly useful in
wireless multihop networks to introduce a hierarchy (e.g. for
routing). Clusters often make use of the concept of domination: nodes
elect a clusterhead and all its neighbors become members of the
cluster \cite{lin97}. However, in some cases two hops may separate
clusterheads so that their interconnection requires gateways.

The Weakly Connected Dominating Set (WCDS) is a well-know structure
often used for network-wide operations such as clustering or
distributing keys in MANET \cite{pathan06}. However, finding a WCDS
with a given cardinality is NP-hard \cite{dunbar97}. Domke et
al. \cite{domke05} extended this result by characterizing graphs
having the same minimum cardinality to form both a WCDS and a
DS. However, the authors focused on particular graphs (e.g. trees with special properties) and did not solve the WCDS problem in any graph. 

Chen et al. \cite{chen02} extended the centralized algorithm for
finding a WCDS of Guha
et al. \cite{guha98} by selecting the best nodes to add in the WCDS
for each round, i.e. the component that forms the WCDS grows at each
step. Dubashi et al. \cite{dubashi05} pruned the edges that belong to
a cycle, i.e. they create a sparser network. Although it forms a
Connected Dominating Set, it cannot directly be used to create a
\rwcds. Alzoubi et al. \cite{alzoubi03} constructed a Maximum
Independent Set, clusterheads being elected based on their depth in a
spanning tree. Thus, this algorithm is close to the \emph{ST}
algorithm presented in the
Section~\ref{section:performance_evaluation}.

In our approach, we build upon the ideas of Chen et al. \cite{chen04}:
they partition the network in zones and each zone executes an
algorithm (a divide-and-conquer approach). However, their algorithm is
greedy and directly applied to each zone. Moreover, they focused on
the original WCDS problem and not on its \rwcds variant. Moreover, we
take into account other performance criteria than the cardinality of
the WCDS, i.e. \emph{network capacity}. Han et al. \cite{han07} adopts
a similar approach of partitioning the network, however the same
remarks as above still hold.


\section{Conclusions and future work}
\label{section:conclusion}

We have presented a divide-and-conquer scheme for computing a
reversible WCDS in a distributed way. By creating a cluster-tree, we
partition the network into clusters with one leader per cluster
solving a \milp formulation to assign roles in its cluster. Although
this approach does not lead to the global optimum, our simulations
show that its performance is very close to a centralized optimal
algorithm.

In the future, we plan to explore new strategies to improve the
convergence of \potatoes. In particular, we can explore redundancy to
simplify the \milp formulation. Moreover, we aim at exploring the
trade-off between optimality and convergence delay: if we pre-assign
some roles, we can reduce complexity along with a negligible impact on
performance.

\section*{Acknowledgments}
This work was partly supported by the European Commission project WIP under contract 2740, and the French Ministry of Research project AIRNET under contract ANR-05-RNRT-012-0.

\bibliographystyle{unsrt}
\bibliography{wcds-potatoes,all}

\end{document}